\documentclass[print]{mn2e}
\usepackage{amsmath}
\usepackage{amssymb}
\usepackage{epsfig}
\usepackage{graphicx}

\title[SGWB from magnetic deformation of magnetars]{Stochastic
gravitational wave background from magnetic deformation of newly
born magnetars}

\author[Cheng et al.]{Quan
Cheng$^{1,2}$\thanks{chengquan1126@126.com}, Yun-Wei Yu$^{2}$
and Xiao-Ping Zheng$^2$\\
$^1$Key Laboratory of Particle Astrophysics, Institute of High
Energy Physics, Chinese Academy of Sciences, Beijing 100049, China\\
$^2$Institute of Astrophysics, Central China Normal University,
Wuhan 430079, China}

\date{Accepted ????. Received ????; in original form 2015 February 5}
\pagerange{\pageref{firstpage}--\pageref{lastpage}} \pubyear{2015}

\begin{document}
\maketitle

\begin{abstract}

Newly born magnetars are promising sources for gravitational wave
(GW) detection due to their ultra-strong magnetic fields and high
spin frequencies. Within the scenario of a growing tilt angle
between the star's spin and magnetic axis, due to the effect of
internal viscosity, we obtain improved estimates of the stochastic
gravitational wave backgrounds (SGWBs) from magnetic deformation of
newly born magnetars. We find that the GW background spectra
contributed by the magnetars with ultra-strong toroidal magnetic
fields of $10^{17}$ G could roughly be divided into four segments.
Most notably, in contrast to the background spectra calculated by
assuming constant tilt angles $\chi=\pi/2$, the background radiation
above 1000 Hz are seriously suppressed. However, the background
radiation at the frequency band $\sim100-1000$ Hz are moderately
enhanced, depending on the strengths of the dipole magnetic fields.
We suggest that if all newly born magnetars indeed have toroidal
magnetic fields of $10^{17}$ G, the produced SGWBs should show sharp
variations with the observed frequency at several tens to about 100
hertz. If these features could be observed through sophisticated
detection of the SGWB using the proposed Einstein Telescope, it will
provide us a direct evidence of the tilt angle evolutions and
further some deep understandings about the properties of newly born
magnetars.

\end{abstract}

\begin{keywords}
gravitational waves --- stars:magnetars --- stars:magnetic field
\end{keywords}

\section{INTRODUCTION}

Neutron stars (NSs) are one of the most promising targets for
gravitational wave (GW) detection with the ground based GW
interferometers such as LIGO, Virgo, GEO600, advanced LIGO (aLIGO)
and the proposed Einstein Telescope (ET; see Sathyaprakash \& Schutz
2009 for a review). In general, GWs can be radiated from NSs because
of many different reasons, e.g., magnetic deformation of the NSs
(Bonazzola \& Gourgoulhon 1996; Regimbau \& de Freitas Pacheco 2001,
2006a; Stella et al. 2005; Dall'Osso et al. 2009; Marassi et al.
2011), dynamical bar-mode instability (Lai \& Shapiro 1995), {\it
r-}mode instabilities due to GW radiation back-reaction (Andersson
1998; Friedman \& Morsink 1998; Zhu et al. 2011), oscillations or
flows excited by glitches (Abadie et al. 2011; Stopnitzky \& Profumo
2014), collapse induced by phase transition to quark matter
(Marranghello et al. 2002; Sigl 2006), and coalescences of compact
binaries (i.e., NS--NS, NS--white dwarf, or NS--black hole;
Schneider et al. 2001; Regimbau \& de Freitas Pacheco 2006b;
Regimbau \& Chauvineau 2007; Zhu et al. 2013). For a newly born NS
of interest here, which if spins initially at the mass-shedding
limit with a millisecond period and has an ultra-strong magnetic
field, it could generate a prompt GW signal due to the bar- and
r-mode instabilities and a continuous GW radiation due to a
magnetically-induced quadrupole ellipticity. Such an ellipticity is
mainly determined by the equation of state (EOS) of the NS, the
magnetic configuration in the stellar interior and, most
importantly, the magnetic energy (Haskell et al. 2008; Dall'Osso et
al. 2009).

It is difficult, usually impossible, to detect GW radiation from
magnetic deformation of the Galactic NSs. For canonical NSs such as
the Crab pulsar and the central compact object harbouring in
Cassiopeia A, their present magnetic fields are obviously too low to
sustain a sufficiently high ellipticity. Direct searches for GWs
from nearby pulsars has put an upper limit of $\epsilon\lesssim
7\times10^{-8}$ at 95\% confidence level for the pulsars (Abbott et
al. 2010). Meanwhile, for the Galactic magnetars whose magnetic
fields could be high enough, their long spin periods ($P\sim2-12$ s;
Olausen \& Kaspi 2014) make them be below the optimal observational
bands of LIGO, Virgo and aLIGO. Nevertheless, fortunately, recent
observations to superluminous supernovae (SLSNe) and gamma-ray
bursts (GRBs) could open a completely new window allowing us to
search the GW signals from some extragalactic newly born NSs. The
fittings to a remarkable number of SLSN light curves (Kasen \&
Bildsten 2010; Inserra et al. 2013) and GRB X-ray afterglows owning
a shallow decay/plateau phase (e.g., Dai \& Lu 1998; Zhang \&
M\'esz\'aros 2001; Yu \& Dai 2007; Yu et al. 2010; Dall'Osso et al.
2011; Rowlinson et al. 2013) robustly suggested that the NSs born
there could be millisecond magentars (i.e., NSs with a high dipole
magnetic field of $\sim10^{14-15}$ G). Such millisecond magnetars
can release a remarkable fraction of their rotational energy in a
sufficiently short period to energize the SLSN and GRB outflows.

On one hand, the millisecond periods of the newly born magnetars
harbouring SLSNe and GRBs can determine the frequencies of their GW
radiation to be within the interval of $\sim10^2-10^3$ Hz, which is
just in the sensitive bands of LIGO and Virgo. On the other hand, in
the sight of dynamo models for magnetic field generation and
amplification, an internal multipole (usually toroidal) magnetic
field much higher than the surface dipole field can be
simultaneously formed due to the differential rotation of the newly
born magnetars (Duncan \& Thompson 1992; Cheng \& Yu 2014). The
equipartition between the differential rotation and the toroidal
field formation would determine the field strength to be on the
order of $\sim10^{16-17}$ G\footnote{An upper limit on the internal
magnetic field can be set by the virial equipartition as $B_{\rm
virial}=1.4\times10^{18}({M_{\rm NS}\over1.4M_\odot})({R_{\rm
NS}\over10^6{\rm cm}})^{-2}$~G (Mastrano et al. 2011).} (Braithwaite
2006). Such values can be supported by the observations to the giant
flare event on 2004 December 27 from SGR 1806-20, which indicates an
internal magnetic field of $\sim 10^{16}$ G in that magnetar (Stella
et al. 2005). The ultra-high internal toroidal magnetic field can
lead to a very high quadrupole ellipticity, allowing the magnetars
to produce strong GW radiation. In fact and specifically, the GW
radiation of a newly born magnetar is sensitive to the tilt angle
between its spin and magnetic axis and, as suggested by Dall'Osso et
al. (2009), the tilt angle is actually time-dependent which is
determined by the competition between the GW radiation and
viscosity. Therefore, for an elaborate calculation of the GW
radiation of newly born magnetars as well as their contributed
stochastic gravitational wave background (SGWB), it is necessary to
take the tilt angle evolution into account, in particular, with an
ultra-high toroidal field.

The SGWB due to magnetic deformation of newly born extragalactic
magnetars has been widely investigated (Regimbau \& de Freitas
Pacheco 2001, 2006a; Marassi et al. 2011), where however the tilt
angle evolutions of magnetars are all neglected which could lead to
some unrealistic consequences. Therefore, in this paper we revisit
the calculation of the SGWB contributed by newly born magnetars by
combining with a consideration of the tilt angle evolution. The
paper is organized as follows. In Sections 2 and 3, the calculations
of the GW spectrum from a single magnetar and the SGWB from all
extragalactic newly born magnetars are introduced, respectively. In
Section 4, we present our numerical results. Conclusion and
discussions are given in Section 5.

\section{Gravitational wave radiation from magnetars}

The ellipticity of magnetars has been investigated widely with
different NS interior structures and different magnetic field
configurations (Bonazzola \& Gourgoulhon 1996; Cutler 2002; Haskell
et al. 2008; Ciolfi et al. 2009, 2010; Mastrano et al. 2011;
Mastrano \& Melatos 2012; Akg{\"u}n et al. 2013; Mastrano et al.
2013; Mastrano et al. 2015). Non-linear numerical simulations showed
that the internal magnetic fields of a magnetar could probably have
a `twisted-torus' configuration (Braithwaite \& Spruit 2004, 2006;
Braithwaite 2009), which is more complicated than the
usually-assumed purely poloidal or toroidal structure. However, in
any case, the toroidal field component is usually found to be the
dominated one. Therefore, in this paper we simply take a pure
toroidal magnetic field in the magnetar interior. In this case, the
quadrupole ellipticity of the magnetar can be estimated as
$\epsilon_{\rm B}=-1.6\times10^{-4}({\bar B}_{\rm t}/10^{16}~{\rm
G})^2$, where ${\bar B}_{\rm t}$ is volume-averaged strength of the
toroidal field (Cutler 2002) and the minus sign indicates that the
magnetar has a prolate shape.

We follow the same procedure as Marassi et al. (2011) in deriving
the GW energy spectrum emitted by a single newly born magnetar and
the SGWB contributed by an ensemble of such magnetars. First, with a
quadrupolar magnetic ellipticity $\epsilon_{\rm B}$ and tilt angle
$\chi$, the rate of energy release from a newly born magnetar via GW
radiation can be written as
\begin{eqnarray}
\frac{dE_{\rm gw}}{dt}=\frac{2G}{5c^5}\epsilon_{\rm
B}^2I^2\omega^6\sin^2\chi(16\sin^2\chi+\cos^2\chi) \label{dEdf1},
\end{eqnarray}
where $I$ is the moment of inertia and $\omega$ the angular
frequency. More specifically, the energy represented by the first
and second term on the right side of the above equation would be
mainly emitted into the GW frequency bands of $\omega/2\pi<\nu_{\rm
e}\leqslant\omega/\pi$ and $\nu_{\rm e}\leqslant\omega/2\pi$,
respectively. Therefore, the spectrum of the GW radiation can be
approximated by
\begin{eqnarray}
\frac{dE_{\rm gw}}{d\nu_{\rm e}}&=&\dot{E}_{\rm
gw}\left|\frac{d\nu_{\rm e}}{dt}\right|^{-1}\nonumber\\
&=&{32\pi G\over5c^5}\epsilon_{\rm
B}^2I^2\omega^6\left|\dot{\omega}^{-1}\right|\sin^4\chi,~{\rm
for~}{\omega\over2\pi}<\nu_{\rm e}\leqslant{\omega\over\pi}\label{dEdf1},\\
&=&\frac{4\pi G}{5c^5}\epsilon_{\rm
B}^2I^2\omega^6\left|\dot{\omega}^{-1}\right|\sin^2\chi\cos^2\chi,~{\rm
for~}\nu_{\rm e}\leqslant{\omega\over2\pi}\label{dEdf2}.
\end{eqnarray}
Due to the GW radiation and magnetic dipole radiation (MDR), the
spin evolution of the magnetar can be determined by
\begin{eqnarray}
\dot{\omega}=-\frac{2G\epsilon_{\rm B}^2I\omega^5}{5c^5}
\sin^2\chi(16\sin^2\chi+\cos^2\chi)-\frac{B_{\rm
d}^2R^6\omega^3}{6Ic^3}{\rm sin}^2\chi \label{dwdt},
\end{eqnarray}
where $B_{\rm d}$ is the strength of the surface dipole magnetic
field at the magnetic pole and $R$ the radius.

Following Cutler \& Jones (2001) and Dall'Osso et al. (2009), on one
hand, the tilt angle can be decreased to zero due to GW radiation,
no matter which shape the magnetar has. On the other hand, for a
magnetar of prolate shape, the tilt angle is inclined to increase to
$\pi/2$ to minimize its precession energy. Therefore an increase of
the tilt angle can be carried out through viscous damping of the
precession energy with a time-scale (Dall'Osso et al. 2009)
\begin{eqnarray}
\tau_{\rm d}=13.5{\rm ~s~}\frac{\cot^2\chi}{1+3{\rm
cos}^2\chi}\left({E_{\rm B}\over 10^{50}~{\rm
erg}}\right)\left({P\over 1~{\rm
ms}}\right)^2\left({T\over10^{10}~{\rm K}}\right)^{-6}\label{vt},
\end{eqnarray}
where $E_{\rm B}$, $P$, and $T$ represent the total magnetic energy,
spin period, and temperature of the magnetar, respectively. In the
following calculations, an analytical thermal history
$T(t)=10^9\left[t/1~{\rm yr}+(10^9~{\rm K}/T_{\rm
i})^6\right]^{-1/6}$ is taken for modified Urca cooling of the
magnetar (Owen et al. 1998). Then following Dall'Osso et al. (2009),
we have
\begin{eqnarray}
{d\sin\chi\over
dt}={\cos^2\chi\over\sin^2\chi}{\sin\chi\over\tau_{\rm
d}}-\frac{2G}{5c^5}\epsilon_{\rm B}^2I\omega^4\sin\chi\label{dchi}.
\end{eqnarray}
It should be noticed that the damping time-scale presented in
Equation (\ref{vt}) is obtained with the bulk viscosity of neutron
matter. In fact, as the temperature of the magnetar decreases to
$\sim 2\times10^9$ K (Page et al. 2004) at the time of $\sim
5\times10^5$ s, the neutron matter in the interior would enter into
superfluid phase and thus the bulk viscosity disappears.
Nevertheless, for temperatures lower than $\sim10^9$ K (Chamel \&
Haensel 2008), a solid stellar crust can be formed to contribute a
new damping effect due to the coupling between the core and crust
(Alpar \& Sauls 1988). The corresponding damping time-scale reads
$\tau_{\rm cc}=nP/\epsilon_{\rm B}$, where $n$ represents the number
of precession cycles (Jones 1976; Cutler 2002). It is difficult to
determine the value of $n$ in theory. For relatively slow rotations,
$n$ is estimated to be $n\sim10^{2-4}$ (Alpar \& Sauls 1988; Cutler
2002). In any case, for $\epsilon_{\rm B}\sim0.01$, it could be
reasonable to assume that tilt angle can be increased to $\pi/2$
immediately along with the crust formation.

\begin{figure}
\resizebox{\hsize}{!}{\includegraphics{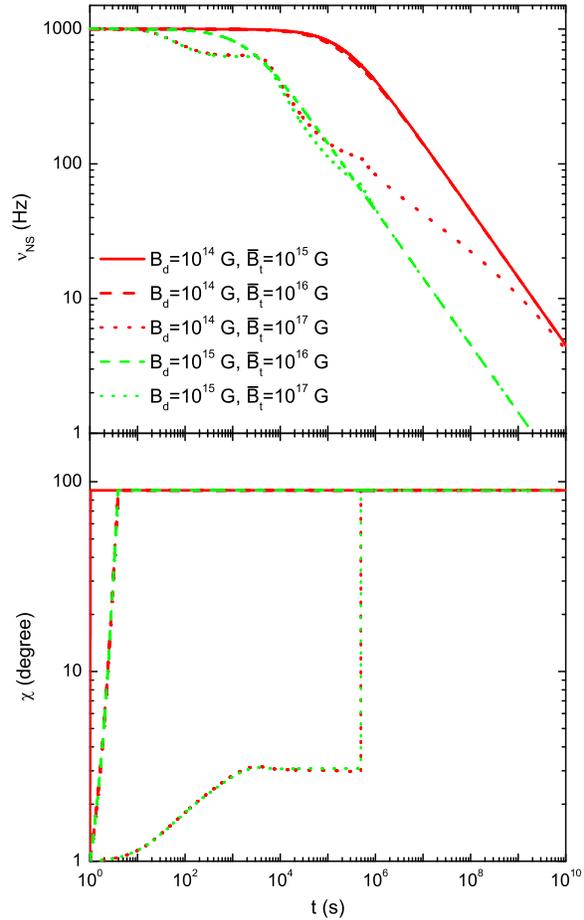}}
\caption{Evolutions of the spin frequency $\nu_{\rm NS}$ (upper
panel) and tilt angle $\chi$ (lower panel) of newly born magnetars.
The strengths of the surface dipole and the internal toroidal
magnetic fields are shown in the legend. The initial conditions for
the magnetars are $P_{\rm i}=1$ ms, $\chi_{\rm i}=1^\circ$, and
$T_{\rm i}=10^{10}$ K. Other quantities for the magnetars are taken
as $M=1.4M_\odot$, $R=10^6$ cm, and $I=10^{45}$ ${\rm g}~{\rm
cm}^2$.}\label{Fig1}
\end{figure}

From the above equations, we can calculate the evolutions of the
spin frequency $\nu_{\rm NS}(=\omega/2\pi)$ and the tilt angle
$\chi$ of magnetars with different parameter values, as presented in
Fig. \ref{Fig1}. In principle, the spin-down of the magnetars is
controlled by both MDR and GW radiation. To be specific, for ${\bar
B}_{\rm t}\ll 10^{17}$ G the MDR braking can always play a
dominative role in the spin-down history, whereas for ${\bar B}_{\rm
t}=10^{17}$ G the early spin evolution would be completely changed
by the strong GW radiation. These results confirm the conclusions of
Dall'Osso et al (2015). Of more interests, in agreement with
Dall'Osso et al. (2009), our results show that the tilt angles of
the magnetars for ${\bar B}_{\rm t}\ll10^{17}$ G can rapidly grow to
$\pi/2$ only in a few seconds, because of the short viscous damping
time-scale for a millisecond spin period. For ${\bar B}_{\rm
t}=10^{17}$ G, even though the viscous damping is suppressed by the
strong GW radiation, the tilt angle can still be increased to
$\pi/2$ at about $5\times10^5$ s due to the curst formation and the
consequent core-crust coupling. In other words, for newly born
millisecond magnetars, the orthogonal configurations with
$\chi=\pi/2$ can be built quickly in all situations, which are the
most benefit for their GW radiation.

\begin{figure}
\resizebox{\hsize}{!}{\includegraphics{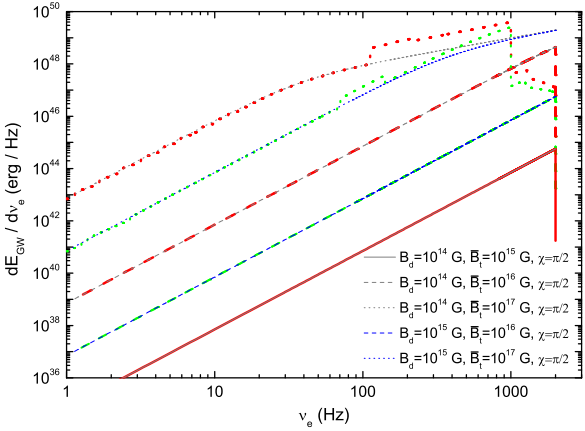}} \caption{The GW
energy spectra emitted by newly born magnetars with different
strengths of the dipole and toroidal magnetic fields. The red and
green lines represent the results calculated by involving the tilt
angle evolutions. The dipole and toroidal fields are the same as
those in Fig. \ref{Fig1}. The grey and blue lines show the spectra
derived only by assuming constant angles $\chi=\pi/2$, but with the
same strengths of magnetic fields as in the cases of $\chi$
evolution, as shown in the legend.}\label{Fig2}
\end{figure}

With the above stellar evolutions, we show the corresponding GW
radiation spectra in Fig. \ref{Fig2}, where some results for
constant tilt angles $\chi=\pi/2$ are also presented for
comparisons. As shown, for ${\bar B}_{\rm t}\ll 10^{17}$ G, the
influence of the tilt angle evolution on the GW radiation spectrum
is very limited, since the angle changes too quickly. However, for
${\bar B}_{\rm t}=10^{17}$ G, some significant changes in the
spectrum appear, in particular, in the high-frequency domain. To be
specific, according to the contributor of radiation, the GW spectrum
can be roughly divided into four segments. For instance, in the case
of $B_{\rm d}=10^{14}$ G and ${\bar B}_{\rm t}=10^{17}$ G, following
Equations (\ref{dEdf1}) and (\ref{dEdf2}), and defining
$A={32\pi^4G\over5c^5}I^2\epsilon_{\rm B}^2\nu_{\rm e}^3$,
$B={B_{\rm d}^2R^6\over6Ic^3}$, $C={2\pi^2G\over5c^5}I\epsilon_{\rm
B}^2\nu_{\rm e}^2$, we have
\begin{eqnarray} \label{dEdnu}
\frac{dE_{\rm gw}}{d\nu_{\rm e}}=\left\{ \begin{aligned}
          &A{\rm sin}^2\chi[B+C(1+15{\rm sin}^2\chi)]^{-1},\cdots&\\
          &1000<\nu_{\rm e}\leq 2000~{\rm Hz}&\\
          &A{\rm cos}^2\chi[B+4C(1+15{\rm sin}^2\chi)]^{-1},\cdots&\\
          &222\lesssim \nu_{\rm e}\leq 1000~{\rm Hz}&\\
          &A{\rm cos}^2\chi[B+4C(1+15{\rm sin}^2\chi)]^{-1}\cdots&\\
          &+A(B+16C)^{-1},~111\lesssim \nu_{\rm e}\lesssim 222~{\rm Hz}&\\
          &A(B+16C)^{-1},~\nu_{\rm e}\lesssim 111~{\rm Hz}.&\\
                          \end{aligned} \right.
                          \end{eqnarray}
From the above equation we know that the first and fourth segments
of the spectrum are contributed by the radiation at $2\nu_{\rm NS}$.
While the second segment is dominated by the radiation at $\nu_{\rm
NS}$ and the third part contains the contributions from $\nu_{\rm
NS}$ and $2\nu_{\rm NS}$. On the other hand, the GW energy spectrum
calculated for $\chi=\pi/2$ can be expressed as
\begin{eqnarray}
\frac{dE_{\rm gw}}{d\nu_{\rm e}}=A(B+16C)^{-1},~\nu_{\rm
e}\leq2000~{\rm Hz}\label{dEdf3}.
\end{eqnarray}

Dividing the four segments of Equation (\ref{dEdnu}) with
(\ref{dEdf3}), respectively, one can see the GW radiation is reduced
by ${B+16C\over B+C}{\rm sin}^2\chi\simeq16{\rm sin}^2\chi$ (due to
$B\ll C$) at the frequency band $1000<\nu_{\rm e}\leq 2000$ Hz;
enhanced by ${B+16C\over B+4C}{\rm cos}^2\chi\simeq4$ (due to $B\ll
C$) at $222\lesssim \nu_{\rm e}\leq 1000$ Hz; enhanced by
$4.4\lesssim 1+{B+16C\over B+4C}{\rm cos}^2\chi \lesssim 5$ (due to
$B\ll C\rightarrow B\simeq C$) at $111\lesssim \nu_{\rm e}\lesssim
222$ Hz; unchanged at $\nu_{\rm e}\lesssim111$ Hz as compared with
the result derived by assuming $\chi=\pi/2$. These analytic results
accurately account for the enhancement in the GW energy spectrum as
shown in Fig. \ref{Fig2}.

Likewise, in the case of $B_{\rm d}=10^{15}$ G and ${\bar B}_{\rm
t}=10^{17}$ G, the GW energy spectrum could also be divided into
four segments, namely, $1000<\nu_{\rm e}\leq 2000$ Hz, $140\lesssim
\nu_{\rm e}\leq 1000$ Hz, $70\lesssim \nu_{\rm e}\lesssim 140$ Hz
and $\nu_{\rm e}\lesssim 70$ Hz. From high- to low-frequency band
the expression for each segment is the same as Equation
(\ref{dEdnu}). Owing to the rather small tilt angle at early period,
the GW radiation is also suppressed at $1000<\nu_{\rm e}\leq 2000$
Hz. However, as $B_{\rm d}$ increases, much more rotational energy
is released into MDR. Hence, the enhancement in the GW radiation is
not as remarkable as in the previous case. Quantitatively, the GW
energy spectrum is enhanced by $1\leq {B+16C\over B+4C}{\rm
cos}^2\chi\lesssim 3.4$ ($B\lesssim C\rightarrow B\gg C$) and
$1+{B+16C\over B+4C}{\rm cos}^2\chi\approx 2$ ($B\gg C$) at
$140\lesssim \nu_{\rm e}\leq 1000$ Hz and $70\lesssim \nu_{\rm
e}\lesssim 140$ Hz, respectively. Moreover, we confirmed the result
of Marassi et al. (2011), which suggested that at low frequency band
where MDR dominates the spin-down, the GW energy spectrum is reduced
by two order of magnitudes while increasing $B_{\rm d}$ from
$10^{14}$ to $10^{15}$ G.

\section{Stochastic GW background}

Generally, the SGWB is denoted by a dimensionless quantity,
$\Omega_{\rm GW}(\nu_{\rm obs})$, which actually represents how the
GW energy density is distributed with the frequency $\nu_{\rm obs}$
in the observer frame. Following Ferrari et al. (1999), this
quantity takes the form
\begin{eqnarray}
\Omega_{\rm GW}(\nu_{\rm obs})=\frac{1}{c^3\rho_{\rm c}}\nu_{\rm
obs}F_{\nu}(\nu_{\rm obs})
\label{SGWB},
\end{eqnarray}
where $\rho_{\rm c}=3H_0^2/8\pi G$ is the critical energy density
needed to close the universe. The GW flux at the observed frequency
$\nu_{\rm obs}$ is
\begin{eqnarray}
F_{\nu}(\nu_{\rm obs})=\int f_{\nu}(\nu_{\rm obs})dR(z), \label{Fnu}
\end{eqnarray}
where $f_{\nu}(\nu_{\rm obs})$ is the GW energy flux per unit
frequency of a single source, $dR(z)$ is the magnetar formation rate
located between $z$ and $z+dz$. The GW energy flux emitted by an
single magnetar is related to its GW energy spectrum as (Regimbau \&
de Freitas Pacheco 2006a)
\begin{eqnarray}
f_{\nu}(\nu_{\rm obs})=\frac{1}{4\pi d_L^2}\frac{d E_{\rm gw}}{d
\nu_{\rm e}}(1+z), \label{fnu}
\end{eqnarray}
where $d_L$ is the luminosity distance, and the GW frequency in the
source frame has the form $\nu_{\rm e}=\nu_{\rm obs}(1+z)$.

The number of NSs formed per unit time out to redshift $z$ can be
estimated by integrating the cosmic star formation rate (CSFR)
density, $\dot{\rho}_{\ast}(z)$, over the comoving volume element,
and taking into account the restriction associated with the initial
mass function (IMF). Of all the NSs, $8-10\%$ are considered to be
magnetars ($B_{\rm d}\geq 10^{14}$ G) according to the results of
Regimbau \& de Freitas Pacheco (2001) and Popov et al. (2009), who
derived this ratio through population synthesis methods. In this
paper, we assume the ratio to be $\xi=10\%$. As a consequence, the
number of magnetars formed per unit time out to redshift $z$ within
the comoving volume can be written as
\begin{eqnarray}
R(z)=\xi\int_0^z\dot{\rho}_{\ast}(z^{\prime})\frac{d V}{d
z^{\prime}}d z^{\prime}\int_{m_{\rm min}}^{m_{\rm max}} \Phi(m)dm,
\label{Rsfr}
\end{eqnarray}
where $dV/dz$ is the comoving volume element, and $\Phi(m)$ is the
IMF. We take the CSFR density model suggested in Hopkins \& Beacom
(2006), which has refined the previous models up to redshift $z\sim
6$, and derived a parametric fit expression for the CSFR based on
the new measurements of the galaxy luminosity function in the UV and
far-infrared wavelengths. The CSFR density can be expressed as
\begin{eqnarray}
\dot{\rho}_{\ast}(z)=h\frac{0.017+0.13z}{1+(z/3.3)^{5.3}}M_\odot
{\rm yr}^{-1}{\rm Mpc}^{-3},
\end{eqnarray}
where $h=0.7$.

The comoving volume element in Equation (\ref{Rsfr}) takes the
following form (Regimbau \& de Freitas Pacheco 2001, 2006a; Zhu et
al. 2011)
\begin{eqnarray}
\frac{d V}{d z}=4\pi\frac{c}{H_0}\frac{r(z)^2}{E(\Omega,z)},
\label{dVdz}
\end{eqnarray}
where $E(\Omega,z)=\sqrt{\Omega_m(1+z)^3+\Omega_\Lambda}$, and
$r(z)=d_L/(1+z)$ is the comoving distance, ${H_0}$ is the Hubble
constant. We take the $\Lambda$CDM cosmological model with
$H_0=70~{\rm km~s}^{-1}{\rm Mpc}^{-1}$, $\Omega_m=0.3$ and
$\Omega_\Lambda=0.7$. The standard Salpeter IMF
$\Phi(m)=Am^{-(1+x)}$ is adopted with $x=1.35$, where $A$ is a
normalization constant, determined by the relation
$\int_{0.1M_\odot}^{125M_\odot}m\Phi(m)dm=1$. It should be noted
that the mass range of magnetar progenitor is still controversial
(see Ferrario \& Wickramasinghe 2008; Davies et al. 2009). However,
we mainly focus on the effect of the tilt angle evolution on the
resultant SGWB. For simplicity, we take the same mass range $m_{\rm
min}=8M_\odot$, $m_{\rm max}=40M_\odot$ as Marassi et al. (2011) for
the magnetar progenitors.

Combining the Equations (\ref{SGWB}), (\ref{Fnu}), (\ref{fnu}),
(\ref{Rsfr}) and (\ref{dVdz}), one can obtain the SGWB contributed
by the magnetically-induced deformation of an ensemble of newly born
magnetars
\begin{eqnarray}
\Omega_{\rm GW}(\nu_{\rm obs})=\frac{8\pi G\nu_{\rm
obs}}{3H_0^3c^2}\int_0^{z_{\rm
upp}}\frac{\dot{\rho}_{\ast}(z)}{(1+z)E(\Omega,z)}\frac{d E_{\rm gw}}{d \nu_{\rm e}}dz   \nonumber\\
\times\int_{m_{\rm min}}^{m_{\rm max}}\Phi(m)dm. \label{omegaGW}
\end{eqnarray}
The upper limit of the redshift integration is determined by the
maximal GW frequency in the source frame and the maximal redshift
$z_{\ast}$ of the CSFR model. To be specific, $z_{\rm
upp}=$min$(z_{\ast}, \nu_{\rm e,max}/\nu_{\rm obs}-1)$. In deriving
Equation (\ref{omegaGW}), the time-dilation effect has been
involved, so the term $(1+z)$ appears in the denominator.

\section{RESULTS}

\begin{figure}
\resizebox{\hsize}{!}{\includegraphics{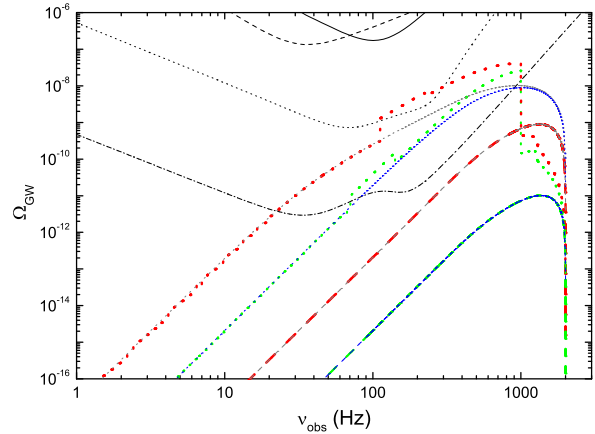}}
\caption{Dimensionless GW energy density $\Omega_{\rm GW}$ versus
the observed frequency $\nu_{\rm obs}$ for newly born magnetars with
different strengths of the dipole and toroidal magnetic fields,
which are the same as in Figs. \ref{Fig1} and \ref{Fig2} (see the
legends in these figures). The red and green lines represent the
SGWBs calculated by involving the tilt angle evolutions. The grey
and blue lines show the SGWBs derived only by assuming $\chi=\pi/2$,
but with the same strengths of magnetic fields as in the cases of
$\chi$ evolution. Please note that the background spectra
contributed by magnetars with $B_{\rm d}=10^{14}$ G and ${\bar
B}_{\rm t}=10^{15}$ G are not shown for both the $\chi$ evolution
case and the $\chi=\pi/2$ case because they are too weak. The
detection thresholds of LIGO (black solid line), VRIGO (black dashed
line), aLIGO (black dotted line), and the proposed ET (black
dash-dotted line), calculated using Equation (136) in Sathyaprakash
\& Schutz (2009) and assuming 1 yr observation time are also
shown.}\label{Fig3}
\end{figure}

In Fig. \ref{Fig3} we plot the dimensionless energy density
$\Omega_{\rm GW}$ versus the observed frequency $\nu_{\rm obs}$ for
newly born magnetars with different strengths of toroidal and dipole
magnetic fields. Since for a magnetar with ${\bar B}_{\rm t}\ll
10^{17}$ G, its tilt angle evolution could not change the emitted GW
energy spectrum, the background spectrum contributed by an ensemble
of such magnetars is also unchanged even their angle evolution is
involved. However, for newly born magnetars with ${\bar B}_{\rm
t}=10^{17}$ G, the background spectra are obviously changed by the
angle evolutions. These magnetars could keep rather small tilt
angles for $\sim5\times10^5$ s, during which they have largely spun
down. As a consequence, the GW radiation from $2\nu_{\rm NS}$ are
seriously suppressed, resulting in the rather weak backgrounds above
1000 Hz. The background spectra show two cutoffs with frequencies at
1000 and 2000 Hz, respectively. Similar cutoffs are also presented
in the spectrum contributed by magnetars with constant tilt angles
$\pi/60$ (Marassi et al. 2011). Moreover, the background spectra
contributed by newly born magnetars with ${\bar B}_{\rm t}=10^{17}$
G can also be divided into four segments.

Specifically, for $B_{\rm d}=10^{14}$ G and ${\bar B}_{\rm
t}=10^{17}$ G, the first segment of the spectrum is at the frequency
band $1000<\nu_{\rm obs}\leq 2000$ Hz, the dimensionless energy
density $\Omega_{\rm GW}$ is reduced by a factor $16{\rm sin}^2\chi$
compared with result of $\chi=\pi/2$. The evolution of $\Omega_{\rm
GW}$ with $\nu_{\rm obs}$ above 1000 Hz is mainly due to the tilt
angle evolution at early periods. Initially, the tiny angles lead to
rather small $\Omega_{\rm GW}$ around 2000 Hz. Subsequently, as the
tilt angles increase and then keep almost constant values of
$3^\circ$, $\Omega_{\rm GW}$ increases and forms a plateau at
$1000<\nu_{\rm obs}\lesssim1250$ Hz. However, such an evolution
behaviour of the background spectrum is unlikely to be detected
currently. The second and third segments of the background spectrum
are at the band $222\lesssim\nu_{\rm obs}\leq 1000$ Hz and
$111\lesssim\nu_{\rm obs}\lesssim 222$ Hz, respectively. Quite the
contrary, by involving the tilt angle evolution, $\Omega_{\rm GW}$
at the two bands are respectively increased by about 4, and $4.4-5$
times, following the analysis in Section 2. The enhanced SGWB could
be detected by aLIGO at a few hecto-hertz. The main reason that
leads to the enhancement is though the tiny tilt angle can suppress
the GW radiation at $2\nu_{\rm NS}$ of a single magnetar, more
rotational energy is released into GW with emitted frequency at
$\nu_{\rm NS}$ instead. The fourth segment is at $\nu_{\rm
obs}\lesssim 111$ Hz, which has no differences with that derived by
assuming $\chi=\pi/2$.

The background spectrum produced by magnetars with $B_{\rm
d}=10^{15}$ G and ${\bar B}_{\rm t}=10^{17}$ G also contains four
segments, which are respectively $1000<\nu_{\rm obs}\leq 2000$ Hz,
$140\lesssim \nu_{\rm obs}\leq 1000$ Hz, $70\lesssim \nu_{\rm
obs}\lesssim 140$ Hz, and $\nu_{\rm obs}\lesssim 70$ Hz. Compared
with the result of $\chi=\pi/2$, the first segment is also reduced
by $16{\rm sin}^2\chi$, while the fourth segment keeps the same.
Nevertheless, the second and third parts are enhanced by $\lesssim3$
times and 2 times, respectively. Overall, stronger $B_{\rm d}$ could
lead to much more rotational energy release in the form of MDR,
rather than GW radiation, further result in a relatively weak SGWB.
As shown in Fig. \ref{Fig3}, the background produced by newly born
magnetars with $B_{\rm d}=10^{15}$ G and ${\bar B}_{\rm t}=10^{17}$
G may only be detected by ET. Actually, if the GW radiation braking
is dominative in the early spin evolution of the magnetars, the
resultant background spectrum is relatively flat at a few tens to a
few hundred hertz, because $dE_{\rm gw}/d\nu_{\rm e}\propto \nu_{\rm
e}$ (GW radiation dominated) as compared to $dE_{\rm gw}/d\nu_{\rm
e}\propto \nu_{\rm e}^3$ (MDR dominated) (Marassi et al. 2011).

One can expect that if all newly born magnetars indeed have
ultra-strong toroidal magnetic fields of $10^{17}$ G, the resultant
dimensionless GW energy densities would show sharp variations with
the observed frequency at several tens to about 100 hertz. We
suggest that if these features could be observed through precise
detection of the SGWB by ET, it will provide us a key evidence about
the early evolution of tilt angles. In addition, this would provide
us some significant information about the properties of newly born
magnetars, such as their braking, cooling mechanisms and the
viscosity of nuclear matter. Of course, the current non-detection of
the SGWB by aLIGO mainly has the following reasons: (i) not all
newly born magnetars have ultra-strong toroidal magnetic fields of
$10^{17}$ G; (ii) of all the newly born NSs, magnetars may have a
ratio less than 10\%; (iii) if all newly born magnetars have ${\bar
B}_{\rm t}=10^{17}$ G, then most of them should have $B_{\rm
d}=10^{15}$ G at least after their births.

\section{CONCLUSION AND DISCUSSIONS}

We have revisited the SGWB contributed by the magnetically-induced
deformation of newly born magnetars involving the tilt angle
evolutions based on the work of Dall'Osso et al. (2009). We find
that for the magnetar with toroidal magnetic fields ${\bar B}_{\rm
t}\ll10^{17}$ G, its tilt angle evolution have no influence on the
emitted GW energy spectrum. Hence, the SGWB from an ensemble of such
magnetars is also unaffected by the angle evolution. However, for
magnetars with ${\bar B}_{\rm t}=10^{17}$ G, their tilt angle
evolutions could obviously change the GW energy spectra and the
background spectra, which both could roughly be divided into four
segments. Specifically, the background radiation above 1000 Hz are
suppressed by a factor of $16{\rm sin}^2\chi$. The intermediate two
segments range from $\sim100-1000$ Hz, the background radiation of
which are enhanced by about four times, and $4.4-5$ times for
$B_{\rm d}=10^{14}$ G ($\lesssim3$ times, and 2 times for $B_{\rm
d}=10^{15}$ G), respectively. Since more rotational energy of the
magnetars are released into MDR for higher $B_{\rm d}$, leading to
the relatively weak enhancement in the GW background radiation. The
last segments are at $\nu_{\rm obs}\lesssim 100$ Hz, the spectra of
which are the same as the results of $\chi=\pi/2$. As a consequence,
involving the tilt angle evolutions could result in some more
stronger and detectable SGWBs, if the newly born magnetars indeed
have ultra-strong toroidal magnetic fields. We expect that the
evidence of tilt angle evolutions of newly born magnetars may
possibly be found in their SGWBs. Moreover, sophisticated detection
of the SGWB at the band below a few hundred hertz may put some
constrains on the properties of these magnetars.

Finally, two aspects that may affect the tilt angle evolution should
be considered in detail in future work. As we know, the bulk
viscosity of quark matter is much larger than that of neutrons,
protons and electrons nuclear matter. As a result, if the magnetars
are quark stars or even hybrid stars, their tilt angle evolutions
should be different from those of classical NSs considered in this
paper. The resultant SGWBs are probably also different. On the other
hand, the ellipticity and magnetic energy of the magnetar with
`twisted-torus' magnetic field configuration are different from that
derived based on the toroidal-dominated configuration (Cutler 2002;
Haskell et al. 2008; Ciolfi et al. 2009, 2010; Mastrano et al. 2011;
Mastrano \& Melatos 2012; Akg{\"u}n et al. 2013; Mastrano et al.
2013; Mastrano et al. 2015). Hence, if the internal fields of
magnetars indeed have a `twisted-torus' shape, we may expect
different tilt angle evolutions, which further result in various
SGWBs.

\noindent\textbf{ACKNOWLEDGEMENTS}

\noindent We thank the anonymous referee for helpful discussions in
improving this paper. This work is supported by the National Natural
Science Foundation of China (grant nos 11133002, 11473008 and
11178001), the funding for the Authors of National Excellent
Doctoral Dissertations of China (grant no. 201225), and the Program
for New Century Excellent Talents in University (grant no.
NCET-13-0822).


\begin{thebibliography}{}

\bibitem{Abadie:2011}Abadie J., et al. 2011, Phys. Rev. D, 83,
042001

\bibitem{Abbott:2010}Abbott B. P., et al. 2010, ApJ, 713, 671

\bibitem{Akgun:2013}Akg{\"u}n T., Reisenegger A., Mastrano A., Marchant P., 2013, MNRAS, 433, 2445

\bibitem{Alpar:1988}Alpar A., Sauls J. A., 1988, ApJ, 327, 723

\bibitem{Andersson:1998}Andersson N., 1998, ApJ, 502, 708

\bibitem{Bonazzola:1996}Bonazzola S., Gourgoulhon E., 1996, A\&A,
312, 675

\bibitem{Braithwaite:2006a}Braithwaite J., 2006, A\&A, 449, 451

\bibitem{Braithwaite:2009}Braithwaite J., 2009, MNRAS, 397, 763

\bibitem{Braithwaite:2004}Braithwaite J., Spruit H. C., 2004, Nature, 431, 819

\bibitem{Braithwaite:2006b}Braithwaite J., Spruit H. C., 2006, A\&A, 450, 1097

\bibitem{Chamel:2008}Chamel N., Haensel P., 2008, Living Reviews in Relativity, 11, 10

\bibitem{Cheng:2014}Cheng Q., Yu Y. W., 2014, ApJL, 786, L13

\bibitem{Ciolfi:2010}Ciolfi R., Ferrari V., Gualtieri L., 2010,
MNRAS, 406, 2540

\bibitem{Ciolfi:2009}Ciolfi R., Ferrari V., Gualtieri L., Pons J.
A., 2009, MNRAS, 397, 913

\bibitem{Cutler:2002}Cutler C., 2002, Phys. Rev. D, 66, 084025

\bibitem{Cutler:2001}Cutler C., Jones D. I., 2001, Phys. Rev. D, 63, 024002

\bibitem{Dai:1998}Dai Z. G., Lu T., 1998, Phys. Rev. Lett., 81, 4301

\bibitem{Dall:2015}Dall'Osso S., Giacomazzo B., Perna R., Stella L., 2015, ApJ, 798, 25

\bibitem{Dall:2009}Dall'Osso S., Shore S. N., Stella L., 2009, MNRAS, 398, 1869

\bibitem{Dall:2011}Dall'Osso S., Stratta G., Guetta D., et al. 2011, A\&A, 526, A121

\bibitem{Davies:2009}Davies B., Figer D. F., Kudritzki R. P.,
Trombley C., Kouveliotou C., Wachter S., 2009, ApJ, 707, 844

\bibitem{Duncan:1992}Duncan, R. C., Thompson, C. 1992, ApJL, 392, L9

\bibitem{Ferrari:1999}Ferrari V., Matarrese S., Schneider R., 1999,
MNRAS, 303, 247

\bibitem{Ferrario:2008}Ferrario L., Wickramasinghe D., 2008, MNRAS,
389, L66

\bibitem{Friedman:1998}Friedman J. L., Morsink S. M., 1998, ApJ, 502, 714

\bibitem{Haskell:2008}Haskell B., Samuelsson L., Glampedakis K., Andersson N., 2008, MNRAS, 385, 531

\bibitem{Hopkins:2006}Hopkins A. M., Beacom J. F., 2006, ApJ, 651,
142

\bibitem{Inserra:2013}Inserra C., et al. 2013, ApJ, 770, 128

\bibitem{Jones:1976}Jones P. B., 1976, Astrophys. Space Sci., 45, 669

\bibitem{Kasen:2010}Kasen D., Bildsten L., 2010, ApJ, 717, 245

\bibitem{Lai:1995}Lai D., Shapiro S. L., 1995, ApJ, 442, 259

\bibitem{Marassi:2011}Marassi S., Ciolfi R., Schneider R., Stella L., Ferrari V., 2011, MNRAS, 411, 2549

\bibitem{Marranghello:2002}Marranghello G. F., Vasconcellos C. Z., de Freitas Pacheco J. A., 2002, Phys. Rev. D, 66, 064020

\bibitem{Mastrano:2013}Mastrano A., Lasky P. D., Melatos A., 2013, MNRAS, 434, 1658

\bibitem{Mastrano:2012}Mastrano A., Melatos A., 2012, MNRAS, 421, 760

\bibitem{Mastrano:2011}Mastrano A., Melatos A., Reisenegger A., Akg\"{u}n T., 2011, MNRAS, 417, 2288

\bibitem{Mastrano:2015}Mastrano A., Suvorov A. G., Melatos A., 2015, MNRAS, 447, 3475

\bibitem{Olausen:2014}Olausen S. A., Kaspi V. M., 2014, ApJS, 212, 6

\bibitem{Owen:1998}Owen B. J., Lindblom L., Cutler C., Schutz B. F., Vecchio A., Andersson N., 1998, Phys. Rev. D, 58, 084020

\bibitem{Page:2004}Page D., Lattimer J. M., Prakash M., Steiner A. W., 2004, ApJS, 155, 623

\bibitem{Popov:2009}Popov S. B., Pons J. A., Miralles J. A., Boldin
P. A., Posselt B., 2009, MNRAS, 401, 2675

\bibitem{Regimbau:2007}Regimbau T., Chauvineau B., 2007, Class. Quantum Grav., 24, 627

\bibitem{Regimbau:2001}Regimbau T., de Freitas Pacheco J. A., 2001, A\&A, 374, 182

\bibitem{Regimbau:2006a}Regimbau T., de Freitas Pacheco J. A., 2006a, A\&A, 447, 1

\bibitem{Regimbau:2006b}Regimbau T., de Freitas Pacheco J. A., 2006b, ApJ, 642, 455

\bibitem{Rowlinson:2013}Rowlinson A., et al. 2013, MNRAS, 430, 1061

\bibitem{Sathyaprakash:2009}Sathyaprakash B. S., Schutz B. F., 2009, Living Reviews in Relativity, 12, 2

\bibitem{Schneider:2001}Schneider R., Ferrari V., Matarrese S., Potergies Zwart S. F., 2001, MNRAS, 324, 797

\bibitem{Sigl:2006}Sigl G., 2006, JCAP, 04, 002

\bibitem{Stella:2005}Stella L., Dall'Osso S., Israel G. L., Vecchio A., 2005, ApJL, 634, L165

\bibitem{Stopnitzky:2014}Stopnitzky E., Profumo S., 2014, ApJ, 787, 114

\bibitem{Yu:2010}Yu Y. W., Cheng K. S., Cao X. F., 2010, ApJ, 715, 477

\bibitem{Yu:2007}Yu Y. W., Dai Z. G., 2007, A\&A, 470, 119

\bibitem{Zhang:2001}Zhang B., M\'esz\'aros P., 2001, ApJL, 552, L35

\bibitem{Zhu:2011}Zhu X. J., Fan X. L., Zhu Z. H., 2011, ApJ, 729, 59

\bibitem{Zhu:2013}Zhu X. J., Howell E. J., Blair D. G., Zhu Z. H., 2013, MNRAS, 431, 882

\end{thebibliography}
\end{document}